\documentclass[preprint,letter,showpacs,preprintnumbers,amsmath,amssymb,floatfix]{revtex4}

\usepackage{graphicx}
\usepackage{dcolumn}
\usepackage{bm}

\begin{document}

\title{Transverse momentum spectra of dileptons at NA60}
\author{Kevin Dusling and Ismail Zahed}
\affiliation{Department of Physics \& Astronomy, State University of New York, Stony Brook, NY 11794-3800, U.S.A.}
\date{\today}  

\begin{abstract}

Recently the NA60 collaboration has reported the transverse mass spectra of dimuons coming from In-In collisions at 158 GeV/A.  
The measured yields display a strong invariant mass dependence not typical of radial flow, suggesting that different sources
contribute in different mass regions. We interpret the dimuon transverse mass spectra from an early thermalized partonic phase
and hadronic phase constrained by the strictures of broken chiral symmetry. Each phase develops a specific transverse momentum
dependence by hydrodynamical expansion. We show that a measurement of the momentum anisotropy at NA60 could provide information 
on the dominant emission source (hadronic or partonic) in the intermediate mass region $1.5 \leq M \leq3.0$ GeV.

\end{abstract}

\maketitle  

\section{Introduction}

The observed dileptons coming from heavy-ion collisions can provide useful information 
on the early stages of the collision since electromagnetic probes do not interact with 
the medium after they are produced.  It is therefore hoped that observed dileptons and 
photons could possibly provide an unambiguous signal for the quark gluon plasma.  
However, in any collision, there is also a substantial contribution of dileptons coming 
from the hadronic phase.  An understanding of the resulting hadronic yields can provide 
crucial information on modifications to electromagnetic spectral functions due to chiral 
symmetry restoration.

The recent NA60 experiment at the CERN SPS has measured the invariant mass spectrum of 
low-mass dimuon pairs \cite{NA60,NA602,NA602a} in In-In collisions.  It was seen that 
a large excess remained after subtracting contributions from expected hadronic (the cocktail) 
decays.  The remaining excess was examined by a number of groups \cite{DZ, TR, RR, RHS, ST} and 
was interpreted by a combination of thermal partonic and hadronic contributions with modifications 
to the spectral function due to finite temperature and baryon density. 

Even though the spectral function directly gives the rate of dilepton production, the 
experimentally observed yield differs due to the fact that the rates must be convoluted 
over the full space-time history of the collision region having widely varying temperature, 
baryon density, chemical potential and flow velocity.  In addition, contributions from an
early  partonic phase must be included.  

The NA60 results for the invariant mass spectra are important for pinning down modifications 
to the spectral function at finite temperature and baryon density but do not provide any 
information on collective transverse flow as the spectra are Lorentz invariant (not 
taking into account the nontrivial acceptance at NA60).  More recently the NA60 collaboration 
has released acceptance corrected transverse mass spectrum \cite{NA603} of the resulting dimuon 
yields in different mass windows.  It was found that the spectrum has a shape atypical of radial 
flow with a significant mass dependence suggesting that the different mass regions are populated 
by different sources.

In the following paper we continue the analysis in~\cite{DZ} and analyze the transverse mass spectra 
of the dimuons produced at NA60 using the same rate equations and hydrodynamic evolution. In section
2 we summarize the pertinent partonic and hadronic rates used. In section 3, we present our transverse
spectra and elliptic flow predictions. Our discussion and conclusion follows in section 4.

\section{Dilepton Rates}

We now summarize the rate equations used for the analysis that was preformed in~\cite{DZ}.  
For the partonic contribution above a critical temperature $T_c \approx 170$ MeV  we use 
the standard leading order q\={q} result for massless quarks:  

\begin{equation}
\label{eq:Bornqqbar}
\frac{dR}{d^4q}=\frac{-\alpha^2}{12\pi^4}\frac{1}
{e^{ q^0/T}-1}\Biglb( N_C 
\sum_{q=u,d,s}e^2_q \Bigrb) \Biglb[ 
1+\frac{2T}{|\Vec{q}|}\ln(\frac{n_+}{n_-}) \Bigrb]
\end{equation}
where $N_C$ is the number of colors, $e_q$ the charge of the quarks, 
and $n_\pm=1/(e^{(q_0\pm|\Vec{q}|)/2T}+1)$.

Below $T_c$ we use the rate equations presented in \cite{paper1, paper2, paper3} 
for a hadronic gas at finite temperature and baryon density which are constrained 
entirely by broken chiral symmetry

\begin{eqnarray}
\frac{dR}{d^4q}=\frac{-\alpha^2}{3\pi^3 q^2}\frac{1}
{1+e^{ q^0/T}}(1+\frac{2m^2_l}{q^2})
(1-\frac{4m^2_l}{q^2})^{1/2} \:\:\:\:\:\:\:\:\:\:\:\:\:\:\:\:\:\:\:\:\:\:\:\:\:\:\:\:\:\:\:\:\:\:\:\:\:\:\:\:\:\:\:\:\:\:\:\:\:\:\:\:\:\:\:\:\:\:\:\:\:\:\:\:\:\:\:\:\:\:\:\:\:\:\:\:\:\: \nonumber\\
\times \Biglb[-3q^2\text{Im} {\bf \Pi}_V(q^2)+\frac{1}{f^2_a}
\int{da}{\bf W}^F_1(q,k)+\int{dN}{\bf W}^F_N(q,p)
\Bigrb] \:\:\:\:\:\:\:\:\:\:\:\:\:\:\:\:\:\:\:\:\:\:\:\:\:\:
\end{eqnarray}

where $da$ and $dN$ are the appropriate phase space factors for mesons and nucleons 
respectively as outlined in~\cite{DZ}.  The term in square brackets is obtained after keeping terms 
to first order in an expansion of the thermal structure function in both meson and nucleon density.  
Through the use of three-flavor chiral reduction formulas the terms ${\bf W}^F_1(q,k) \text{ and } 
{\bf W}^F_N(q,p)$ can be expressed in terms of vacuum vector and axial spectral densities which are measured 
in $e^+e^-$ annihilation, $\tau$ decays and photo-reactions on nucleons and nuclei.

The final contribution comes from the decay of vector mesons after freezeout.  
These dileptons will play a large role in the transverse momentum spectra in the mass region 
of the given meson.  The main effect is to harden the transverse mass spectra as the decaying mesons 
at freezeout are coupled to the transverse flow of the medium and are therefore boosted to higher 
$q_T$.  Since the contribution from the decay of freezeout vector mesons was not included in our first 
analysis~\cite{DZ}, we discuss the rate equation and the integration over the freezeout surface in the 
appendix.  

\section{Results}

\begin{figure}
\includegraphics[scale=.4]{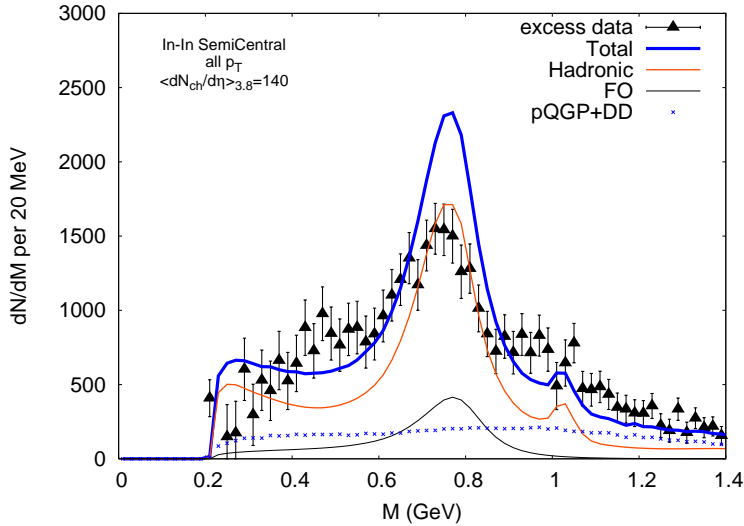}
\caption{Resulting dimuon yields after integration over the space time volume of the collision convoluted 
with a schematic acceptance of the NA60 experiment.  Shown are the contributions from the perturbative 
QGP and charm decays (pQGP + D\={D}), the decay of freezeout $\rho$ mesons (FO), and the thermal 
hadronic yields (Hadronic). }
\label{fig:dM}
\end{figure}

The rates outlined in the previous section were integrated over the space-time history of a full hydrodynamic 
simulation (details discussed in~\cite{DZ}) of the collision region at the CERN SPS collider.  Shown in 
fig.~\ref{fig:dM} are the resulting invariant mass spectra (after a schematic acceptance of the NA60 detector 
was applied) compared to the data measured by the NA60 collaboration.  The only difference from the analysis 
done in~\cite{DZ} is the inclusion of the freezeout contribution as shown in the figure.  The net result is 
the same within \%10 except for a small change in the overall normalization of the rates which can be accounted 
for by a decrease in the fireball freezeout temperature.

\begin{figure}[hbtp]
\includegraphics[scale=.35]{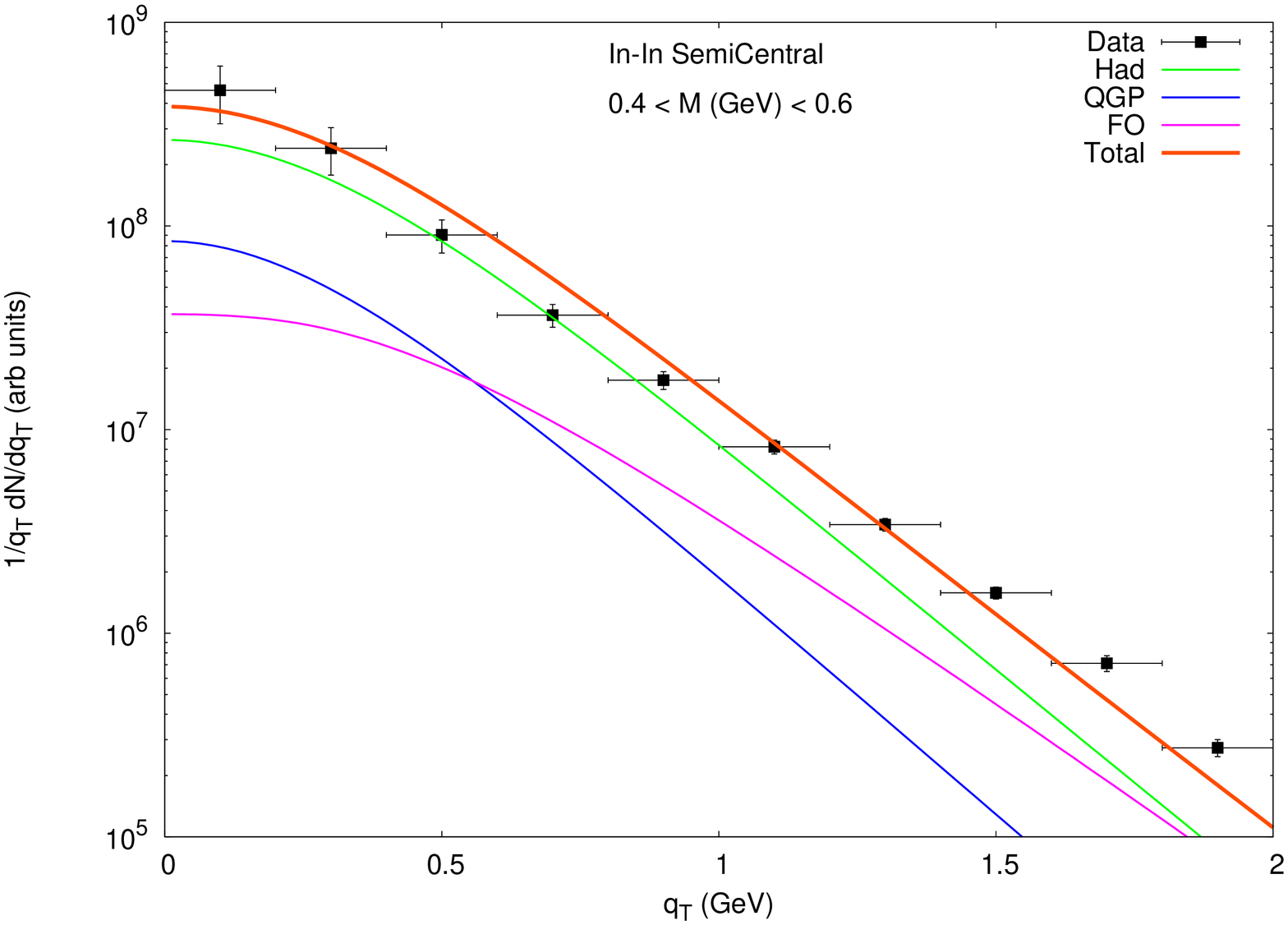}
\\
    \vspace{0.1in}
\includegraphics[scale=.35]{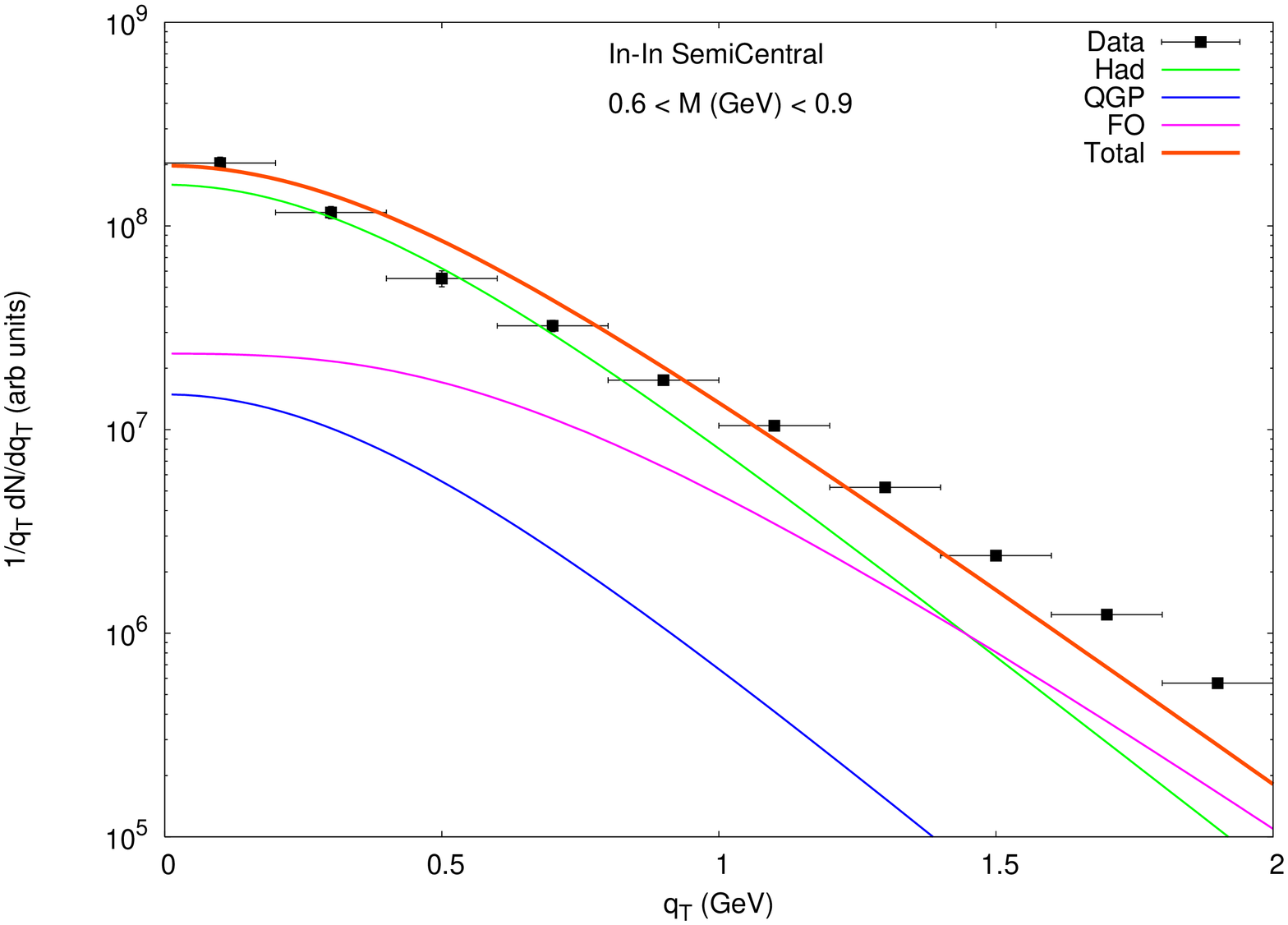}
\\
	\vspace{0.1in}
\includegraphics[scale=.35]{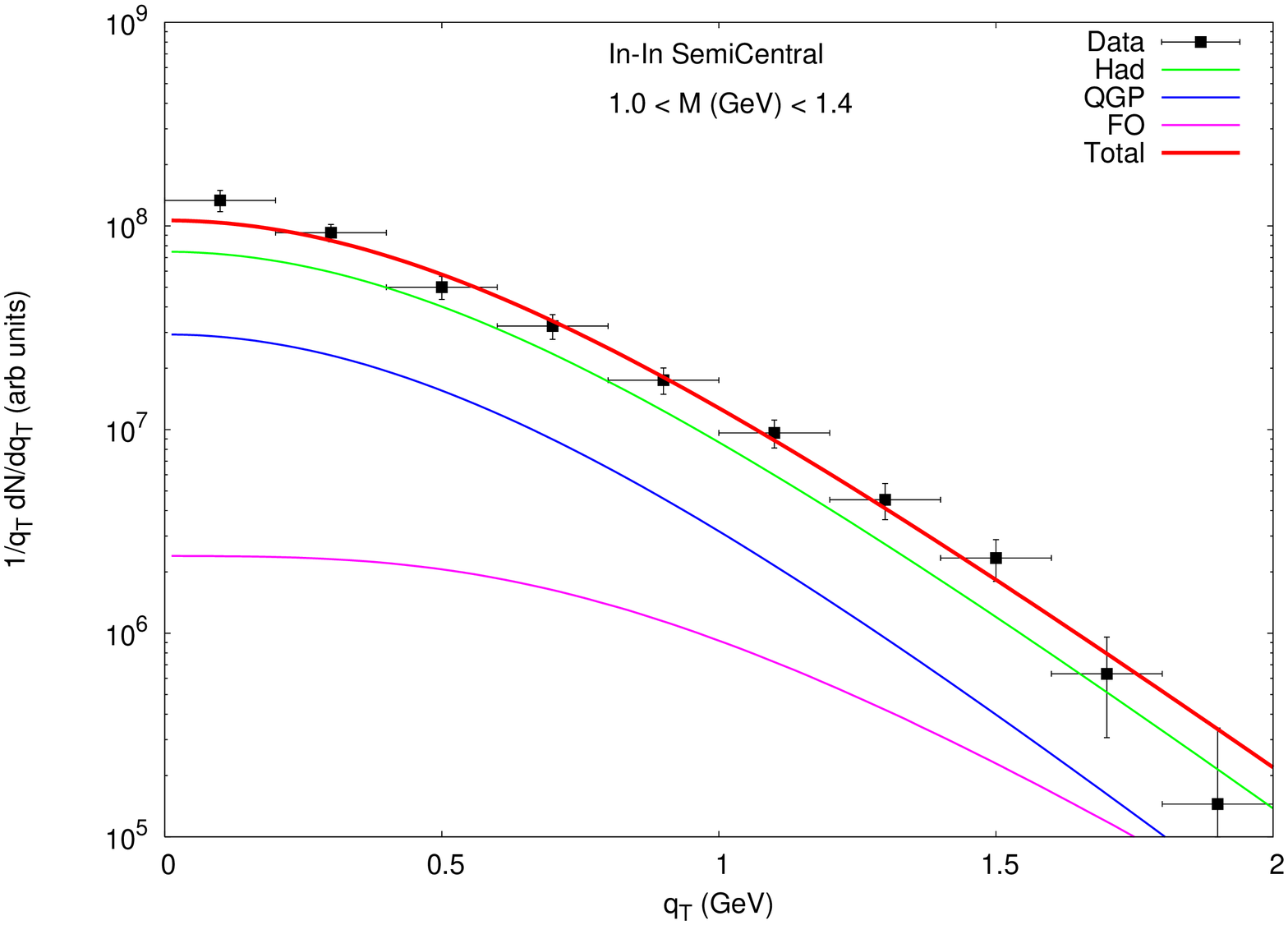}
  \vspace{9pt}

  \caption{(Color Online) Transverse momentum spectra for the three mass bins as measured 
by NA60. From top to bottom: 0.4 $<$ M (GeV) $<$ 0.6, 0.6 $<$ M (GeV) $<$ 0.9, 1.0 $<$ M (GeV) $<$ 1.4.  
The solid lines are our results including contributions from the partonic phase (QGP), 
hadronic phase (Had) and vacuum decays after freezeout (FO).}
  \label{fig:result}
\end{figure}

The data points in fig.~\ref{fig:result} show the acceptance corrected transverse momentum 
spectrum of observed dimuon pairs after subtraction of the known hadronic cocktail 
(omitting the $\rho$) and charm contribution.  The data was divided into three mass windows: 
a low mass region (0.4 $<$ M (GeV) $<$ 0.6) below the $\rho$ peak, a mass region around the 
$\rho$ (0.6 $<$ M (GeV) $<$ 0.9) and also a higher mass region (1.0 $<$ M (GeV) $<$ 1.4).  
The data selection contains $dN_{ch}/d\eta>30$ which corresponds to about $<dN_{ch}/d\eta>\approx140$ 
which we use in the analysis.

The solid lines in fig.~\ref{fig:result} show the results using the rates discussed in the 
prior section.  We show the contribution to the $q_T$ spectrum from three sources: partonic (QGP), 
hadronic (Had) and freeze-out (FO) $\rho$ contribution.  The upper line in each figure shows the total yield.

\begin{figure}
\includegraphics[scale=.65]{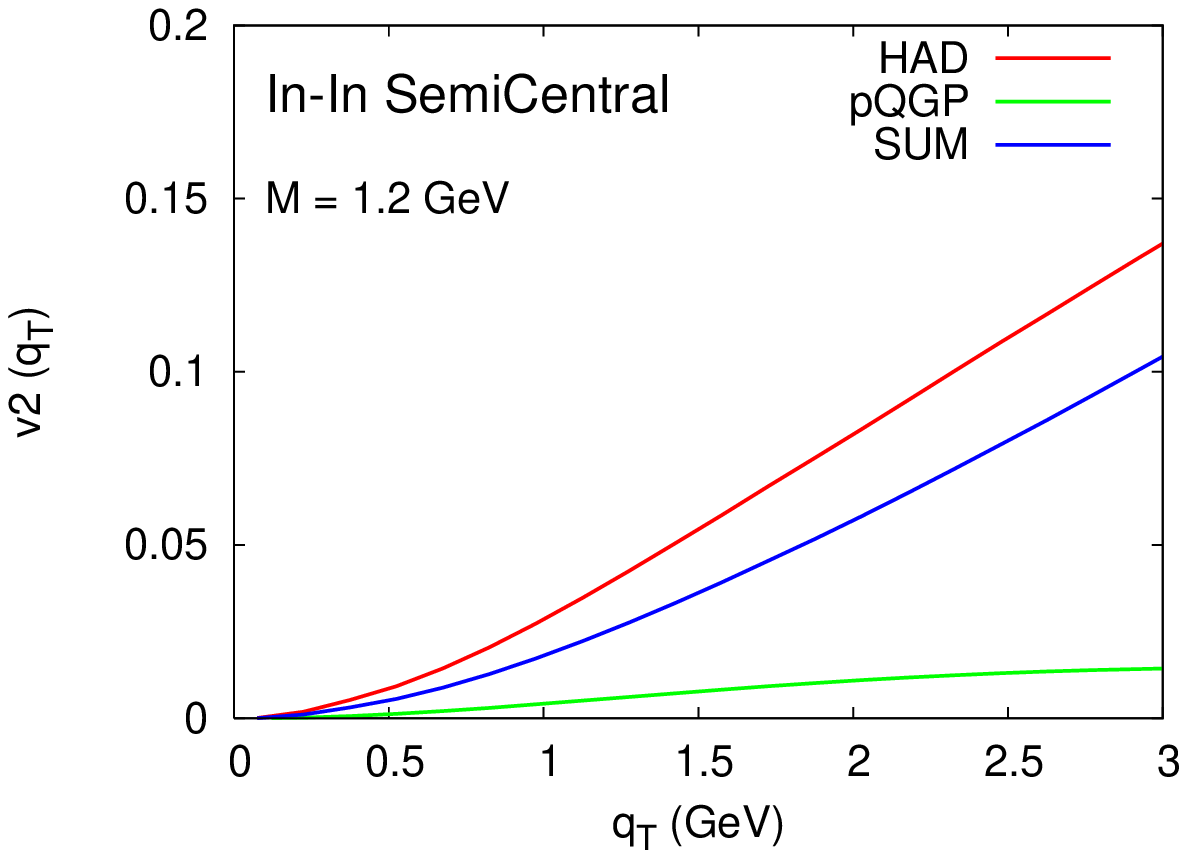}
\caption{(Color Online) Dilepton elliptic flow as a function of $q_T$ for semi-central In-In collisions.}
\label{fig:v2}
\end{figure}

Since a full hydrodynamic simulation of the collision is available we go one step further and calculate 
the elliptic flow of the produced dileptons \cite{Heinz1} as is shown in fig~\ref{fig:v2} where $v2$ is 
a function of the invariant mass, transverse momentum and the impact parameter of the colliding system.  
The elliptic flow is defined as the weighted average of the yields with $\cos(2\phi)$:
\begin{align}
v2(M,q_T)=
{\int d\phi \cos(2\phi)\frac{dN}{dM^2dy q_T dq_T d\phi}}/{\int d\phi\frac{dN}{dM^2dy q_T dq_T d\phi}}
\end{align}
where $\phi$ is the angle between the dilepton momentum ($\vec{q}$) and the corresponding fluid element's velocity.

\section{Discussion and Conclusion}

Our rates are overall consistent with the data in all mass regions.  There is the possibility of some enhancement in 
the high $q_T$ regions in both the low mass and medium mass regions.  This could possibly be accounted for by fine 
tuning the freezeout temperature thereby increasing the freezeout $\rho$ contribution and thus hardening the spectra in 
these mass regions. The drawback however, is that our yield around the rho is already about 30\% above the measured
yield. There is also some discrepancy in the low $q_T$ regions, but as was pointed out in~\cite{RR06}, one should not 
expect to explain this region by a thermal description only.

Based on the data, we feel that overall we have good control over the hydrodynamic evolution of both the temperature 
and flow velocity as well as the different source contributions (hadronic or partonic) in different mass regions.  
Of course with additional fine-tuning of fireball parameters (freeze-out temperature, etc.) it is possible to achieve
better agreement, but we don't feel this is worth doing since the data is integrated over a wide range of centralities.

There has also been discussion whether the mass region above 1 GeV is dominated by partonic or hadronic radiation.  
Even though information on the source can be extracted from $q_T$ spectra by looking at the effective temperature of 
the emission region as was done in~\cite{RR06} more information could be obtained if it was possible to measure the 
dilepton elliptic flow in various mass windows.  Since the thermal partonic radiation is emitted early on from the 
collision region, one does not expect a large momentum anisotropy to have developed and therefore one would expect a 
much lower elliptic flow from dileptons coming from this region.  On the other hand, hadronic radiation is emitted 
much later on and any momentum anisotropy would also be carried by the emitted dileptons.
  
In our rate calculations we find that in the $1.0 < $ M (GeV) $ < 1.5$ region the dilepton contribution comes from 
both the hadronic and partonic phases.  The hadronic yields in this mass region comes from contributions of both 
the axial spectral density which enters the rates from the chiral reduction formula due to chiral symmetry restoration 
by matter and also from the high energy tail of the vector spectral weight, part of which is from 4-$\pi$ annihilation.  
In fig.~\ref{fig:v2} we show our calculated $v2$ from the partonic and hadronic yields.  We also show the total $v2$, 
which is a weighted average of the two sources.  One can see that depending upon the dominant emission source that 
qualitatively different elliptic flow develops at larger $q_T$.

To summarize:
We have interpreted the dimuon transverse mass spectra as measured by the NA60 collaboration as coming from an 
early thermalized partonic phase as well as a hadronic phase constrained entirely by broken chiral symmetry, both 
of which develop specific transverse momentum dependence following the underlying hydrodynamic flow.  After convolution 
of the rates over a hydrodynamic simulation of the conditions at the CERN SPS experiment we find good agreement 
between the resulting yields and the measured data.  We show that a measurement of the momentum anisotropy at NA60 
could provide information on the dominant emission source (hadronic or partonic) in the intermediate mass region 
$1.5 \leq M \leq3.0$ GeV.
\\

{\bf Acknowledgments.}
\\
We would like to thank Sanja Damjanovic, Edward Shuryak and Hans Specht for useful discussions.  This work was partially 
supported by the US-DOE grants DE-FG02-88ER40388 and DE-FG03-97ER4014.

\appendix

\section{Dileptons after freeze-out}

We discuss how the dilepton rates from vector mesons decaying after freezeout was evaluated in the above result.  
A number of works \cite{Kam, Heinz, Ruus} discuss this topic but it is presented here as well for completeness.

The contribution to the dilepton spectrum from decaying vector mesons at freezeout is given by:
\begin{align}
\frac{dN}{dM^2\text{ }d^3q/q^0}=\sum_V \frac{1}{\Gamma^{tot}_V} \frac{d\Gamma_{V\to\mu^+\mu^-}}{dM^2} E \frac{d^3N_V}{d^3p} \delta^4(p-q)
\label{eq:tot}
\end{align}
In the above equation $p$ is the momentum of the decaying vector meson V, $q=l_1+l_2$ is 
the momentum of the dilepton pair and $M=\sqrt{l_1^2+l_2^2}$ is the invariant mass of the dilepton pair.  
The quantity $E\frac{d^3N_V}{d^3p}$ is the number of vector mesons with momentum $\vec{p}$ and energy E 
at freezeout and can be evaluated using the Cooper-Frye \cite{cf} formula
\begin{align}
E\frac{d^3N}{d^3p}=\frac{g}{(2\pi)^3}\int_{\sigma}f(q_0,T)p^\mu d\sigma_\mu
\label{eq:cf}
\end{align}
The Cooper-Frye formula requires integration over $\sigma$, which is the four-dimensional space-time boundary.  
We show how this integration was done for a boost invariant, azimuthally symmetric expansion.

The quantity $d\sigma_\mu$ is the normal vector to the surface of the four-dimensional space-time volume 
set in our case by the freeze-out temperature $T_{f.o.}$ and is given by
\begin{align}
d^3\sigma_\mu \equiv -\epsilon_{\mu\nu\lambda\rho}\frac{\partial\sigma^\nu}{\partial u}
\frac{\partial\sigma^\lambda}{\partial v}\frac{\partial\sigma^\rho}{\partial w}du\text{ }dv\text{ }dw
\label{eq:var}
\end{align}
where u,v,w are three independent coordinates used to parameterize the hypersurface.
Following \cite{bjorken} the four volume of the collision region is expressed as 

\begin{align}
\sigma^\mu(\tau,r,\phi,\eta)=[\tau\cosh\eta,r\cos\phi,r\sin\phi,\tau\sinh\eta]
\end{align}
where $\eta=\frac{1}{2}\ln\frac{t+z}{t-z}$ is the spatial rapidity and $\tau=\sqrt{t^2-z^2}$ 
is the proper time.  Also for a boost invariant expansion, the vector meson's four momentum can be expressed as:
\begin{align}
p^\mu=[ m_\perp \cosh y, p_\perp \cos\phi^p, p_\perp \sin\phi^p, m_\perp \sinh y ]
\end{align}
where $m_\perp=\sqrt{M^2+p_\perp^2}$ and y is the longitudinal rapidity.

Choosing to parameterize the freezeout hypersurface with coordinates (u,v,w)=(r,$\phi$,$\eta$) the freezeout volume is: 
\begin{align}
\sigma^\mu_f(r,\phi,\eta)=[\tau_f(r)\cosh \eta,r\cos \phi ,r\sin \phi,\tau_f(r)\sinh \eta ]
\end{align}
The subscript f in the above equation refers to the quantity at freezeout.  
Therefore $\sigma^\mu_f$ should be thought of as a parameterization of the space-time 
volume where freezeout occurs, or more precisely in our case the space-time volume having 
temperature $T=T_{f.o.}$.  The fact that the freezeout proper time depends only on the fireball 
radius is due to the assumed boost-invariance and azimuthal symmetry.
Using \ref{eq:var} we obtain
\begin{align}
d^3\sigma_\mu=\biglb( \cosh\eta,-\frac{\partial \tau}{\partial r}\cos\phi,-
\frac{\partial \tau}{\partial r}\sin\phi,-\sinh\eta\bigrb)\tau r dr d\phi d\eta
\end{align}
for the normal vector to the hypersurface.

Using the above expressions for $p^\mu \text{ and } d\sigma_\mu$ and performing the 
integration of Eq.~\ref{eq:cf} over the freezeout hypersurface, the total number of 
dileptons produced from decaying $\rho$ mesons per unit four momentum is given as:
\begin{eqnarray}
\frac{dN}{d^4q}=\int_0^{r_{f.o.}} dr \int_{-\infty}^{+\infty} d\eta 
\int_0^{2\pi} d\phi\text{  } \tau r  \:\:\:\:\:\:\:\:\:\:\:\:\:\:\:\:\:\:\:\:\:\:\:\:\:\:\:\:\:\:\:\:\:\:\:\:\:\:\:\:\:\:\:\:\:\:\:\:\:\:\:\:\:\:\:\:\:\:\:\:\:\:\:\:\:\:\:\:\:\:\:\:\:\:\:\:\:\:\:\:\:\:\:\:\:\:\:\:\:\:\:\:\:\:\:\:\:\:\:\:\:\:\:\:\:\:\:\: \nonumber\\
 \times 
\biglb[ m_\perp\cosh(y-\eta)-\frac{\partial \tau}{\partial r}p_\perp\cos(\phi-\phi^p)\bigrb]
\frac{1}{\Gamma^{tot}_\rho}\frac{d\Gamma_{\rho\to\mu^+\mu^-}}{dM^2}\frac{g}{(2\pi)^2}f(p_\mu u^\mu,T) \:\:\:\:\:\:\:\:\:\:\:\:\:\:\:\:\:\:\:
\end{eqnarray}

where we have chosen to orient the azimuthal angle $\phi$ with the direction of 
the fluid velocity and $f(p_\mu u^\mu,T)$ is the Bose distribution with 
$p_\mu u^\mu=m_\perp \gamma_\perp \cosh(y-\eta)-p_\perp u_\perp \cos(\phi-\phi^p)$.  
The explicit form for the meson spectral function is given using the Breit-Wigner parameterization:

\begin{align}
\frac{d\Gamma_{\rho\to\mu^+\mu^-}}{dM^2}=\Gamma_{\rho\to\mu^+\mu^-}
\frac{1}{\pi}\frac{m_\rho\Gamma_\rho^{tot}}{(M^2-m_\rho^2)^2+(m_\rho\Gamma_\rho^{tot})^2}
\end{align}

\end{document}